\documentclass[12pt,letterpaper]{article}
\usepackage[pdftex]{graphicx,color}
\input{psfig}
\input{epsf}
\usepackage{amsmath,amssymb}
\usepackage[dvips,letterpaper,text={6.5in,9in}]{geometry}
\usepackage{fancyhdr}
\usepackage{verbatim}

\newcommand\ltap{\
  \raise.3ex\hbox{$<$\kern-.75em\lower1ex\hbox{$\sim$}}\ }
\newcommand\gtap{\
  \raise.3ex\hbox{$>$\kern-.75em\lower1ex\hbox{$\sim$}}\ }

\newcommand\simge{\mathrel{%
   \rlap{\raise 0.511ex \hbox{$>$}}{\lower 0.511ex \hbox{$\sim$}}}}
\newcommand\simle{\mathrel{
   \rlap{\raise 0.511ex \hbox{$<$}}{\lower 0.511ex \hbox{$\sim$}}}}

\newcommand{\slashchar}[1]%
        {\kern .25em\raise.18ex\hbox{$/$}\kern-.75em #1}
\def\lsim{\mathrel{\raise.3ex\hbox{$<$\kern-.75em\lower1ex\hbox{$\sim$}}}}
\def\gsim{\mathrel{\raise.3ex\hbox{$>$\kern-.75em\lower1ex\hbox{$\sim$}}}}

\newcommand\CM{{\cal M}}

\newcommand\CO{{\cal O}}

\newcommand\be{\begin{equation}}
\newcommand\ee{\end{equation}}
\newcommand\bea{\begin{eqnarray}}
\newcommand\eea{\end{eqnarray}}
\newcommand\ba{\begin{array}}
\newcommand\ea{\end{array}}
\newcommand\nn{\nonumber}

\newcommand{\half}{\ensuremath{\frac{1}{2}}}
\newcommand{\third}{\ensuremath{\frac{1}{3}}}

\newcommand{\thw}{\ensuremath{\theta_W}}

\newcommand\dagg{\dagger}
\newcommand\ts{\thinspace}
\newcommand\ra{\rightarrow}

\newcommand\gev{{\rm GeV}}
\newcommand\tev{{\rm TeV}}

\newcommand\pb{{\rm pb}}
\newcommand\ipb{{\rm pb}^{-1}}
\newcommand\fb{{\rm fb}}
\newcommand\ifb{{\rm fb}^{-1}}

\newcommand\etmiss{\slashchar{E}_T}

\newcommand\ellm{\ell^-}
\newcommand\ellpm{\ell^\pm}
\newcommand\ellp{\ell^+}

\newcommand\Ltc{\Lambda_{TC}}

\newcommand\tom{\omega_{T}}
\newcommand\tro{\rho_{T}}

\newcommand\ta{a_T}

\newcommand\tapm{a_T^\pm}

\newcommand\taz{a_T^0}
\newcommand\at{a_T}

\newcommand\tropm{\rho_{T}^\pm}

\newcommand\troz{\rho_{T}^0}

\newcommand\tpi{\pi_T}
\newcommand\tpipm{\pi_T^\pm}
\newcommand\tpimp{\pi_T^\mp}
\newcommand\tpip{\pi_T^+}

\newcommand\tpiz{\pi_T^0}
\newcommand\tpipr{\pi_T^{0 \prime}}

\newcommand\jet{{\rm jet}}

\begin{document}
\title{
\vskip -15mm
\begin{flushright}
\vskip -15mm
{\small FERMILAB-Pub-07-202-T\\
BUHEP-07-04\\
arXiv 0706.2339\\}
\vskip 5mm
\end{flushright}
{\Large{\bf Low-Scale Technicolor at the Tevatron and LHC}}\\
} \author{
  {\large Estia Eichten$^{1}$\thanks{eichten@fnal.gov} \, and
    Kenneth Lane$^{2}$\thanks{lane@physics.bu.edu}}\\
  {\large {$^1$}Fermi National Accelerator Laboratory}\\
  {\large P.O.~Box 500, Batavia, Illinois 60510}\\
  {\large {$^2$}Department of Physics, Boston University}\\
  {\large 590 Commonwealth Avenue, Boston, Massachusetts 02215}\\
}
\maketitle

\begin{abstract}
  The Tevatron experiments CDF and D\O\ are close to making definitive
  statements about the technicolor discovery mode $\tro \ra W\tpi$ for
  $M_{\tro} \simle 250\,\gev$ and $M_{\tpi} \simle 150\,\gev$. We propose new
  incisive tests for this mode and searches for others that may be feasible
  at the Tevatron and certainly are at the LHC. The other searches include
  two long discussed, namely, $\tom \ra \gamma \tpi$ and $\ell^+\ell^-$, and
  a new one --- for the $I^G J^{PC} = 1^- 1^{++}$ partner, $\ta$, of the
  $\tro$. Adopting the argument that the technicolor contribution to $S$ is
  reduced if $M_{\ta} \simeq M_{\tro}$, we enumerate important $\ta$ decays
  and estimate production rates at the colliders.

\end{abstract}


\newpage


{\em{1. {\underbar{Introduction}}}} Long ago we pointed out that
technicolor's (TC) walking gauge coupling~\cite{Holdom:1981rm,
  Appelquist:1986an,Yamawaki:1986zg,Akiba:1986rr} strongly indicates that the
energy scale of its bound states is much lower than the roughly $1\,\tev$
expected from the earliest TC studies~\cite{Eichten:1984eu}. Thus, the
lightest technihadrons may well be within reach of the Tevatron collider,
with production rates of several picobarns; they certainly are accessible at
the LHC. Furthermore, they should have striking decay signatures. To review
the arguments~\cite{Lane:1989ej,Eichten:1996dx,Eichten:1997yq}: (1)~The
walking TC gauge coupling probably requires either a large number $N_D$ of
technifermion doublets so that $\Ltc \simeq 250\,\gev/\sqrt{N_D} \simle
100\,\gev$, or two TC scales, one much lower than $250\,\gev$. (2)~Walking
enhances the masses of pseudo-Goldstone technipions, $\tpi$, more than those
of their vector partners, $\tro$ and $\tom$. This probably closes the
vectors' all-$\tpi$ decay channels, leaving only those involving at least one
electroweak (EW) gauge boson (especially longitudinally-polarized
$W_L^{\pm,0}$) and perhaps a $\tpi$. Another striking final state is
$\ell^+\ell^-$.  Technipions are expected to decay via extended technicolor
(ETC) interactions~\cite{Eichten:1979ah} to the heaviest flavors possible,
putting a premium on $b$-tagging. Thus, those accessible at the Tevatron will
appear in vector-meson-dominated Drell-Yan processes such as $\bar q q' \ra
\gamma,Z,W \ra \tro \ra W \tpi \ra \ell^\pm \nu_\ell b \bar q$ and $\tom \ra
\gamma \tpiz \ra \gamma {\overline b}b$. At the LHC, large backgrounds tend
to force one into looking at all-EW boson final states ending up in $e$~ and
$\mu$-generation leptons.

The phenomenology of the lightest $\tpi$, $\tro$ and $\tom$ of low-scale
technicolor --- bound states of the lightest technifermion color-singlet
EW doublet, $(T_U,T_D)$ --- is embodied in the ``Technicolor
Straw-Man Model'' (TCSM)~\cite{Lane:1999uh,Lane:2002sm}. The TCSM's most
important assumptions are: (1) These technihadrons may be treated in
isolation, without significant mixing or other interference from higher-mass
technihadrons. (2) The $\bar T T$ technipions $\Pi_T$ are not mass
eigenstates, but they may be treated as simple two-state mixtures of
$W_L^{\pm,0}$ and mass-eigenstate $\pi_T^{\pm,0}$:
\be\label{eq:pistates}
 \vert\Pi_T\rangle = \sin\chi \ts \vert
W_L\rangle + \cos\chi \ts \vert\tpi\rangle\,.
\ee
Here, $\sin\chi = F_T/246\,\gev$, where $F_T$ is the $\tpi$ decay constant.
In a model with $N_D \gg 1$, $\sin\chi \cong 1/\sqrt{N_D}$; in a two-scale
model, $F_T \ll 246\,\gev$~\cite{Lane:1989ej}. This implies that the
technivectors are very narrow, with decay rates suppressed by some
combination of phase space and powers of $\sin\chi$ and/or EW gauge
couplings. (3) Techni-isospin is a good symmetry.  (4) Finally, something
like topcolor-assisted technicolor~\cite{Hill:1994hp} is needed to keep the
top quark from decaying copiously into $\tpip b$ when $M_{\tpi} \simle
160\,\gev$. Thus, if $\tpip$ is heavier than the top, it does not decay
exclusively to $t \bar b$.

Notwithstanding the isolation assumption above, many higher-mass states are
reasonably expected. In Refs.~\cite{Lane:1993wz,Lane:1994pg} it was argued
that walking TC invalidates the standard QCD-based calculations of the
precision-electroweak $S$-parameter~\cite{Peskin:1990zt,Golden:1990ig,
  Holdom:1990tc,Altarelli:1991fk} because the spectral functions in the
integral for $S$ cannot be saturated by just the lightest $\tro$ and its
axial partner, $\ta$. Something more must make the spectral integrals
converge much slower than they do in QCD; the obvious possibility is a tower
of vector and axial-vector isovector mesons.

The main message of this letter is that the lightest $\ta$ is within reach of
the Tevatron and LHC if the $\tro$ and $\tom$ are, and that $\ta$ decays also
produce unusual signatures to aid their discovery. In fact, it has long been
recognized that the $S$-parameter may be significantly reduced or even made
negative if $\tro$ and $\ta$ pairs in the tower are close in mass and have
comparable couplings to their respective currents; see, e.g.,
Refs.~\cite{Appelquist:1998xf,Knecht:1997ts,Appelquist:1999dq}. More
recently, a number of attempts have been made to model walking TC and
calculate $S$ using the AdS/CFT connection; see, e.g., Ref.~\cite{
  Hirn:2006nt,Hirn:2006wg}. We adopt these papers' suggestion that $M_{a_T}
\simeq M_{\rho_T}$ and explore its phenomenological
consequences.\footnote{The $I=0$ partner $h_T$ of $\ta$ presumably is nearly
  degenerate with it, but it cannot be produced via the VMD process at hadron
  and lepton colliders.} For $M_{a_T} \simge M_{\tro} = 200$--$250\,\gev$,
its discovery should be possible at the Tevatron. We expect there would be
little trouble discovering and studying $\ta$ at the LHC. Preliminary studies
for all these technihadrons at the LHC appear in Refs.~\cite{Kreuzer:2006sd,
  Kreuzer:2007zz,Brooijmans:2008se,LSTCazuelos}.

Our second motivation is the appearance of interesting anomalies in searches
for $\tro \ra W^\pm \tpi \ra \ellpm \etmiss b\,\jet$ by the D\O\ and CDF
Collaborations. In 2006 the D\O\ Collaboration presented two analyses of a
$388\,\ipb$ data set, one cut-based, the other using a neural net
(NN)~\cite{Abazov:2006iq}. While it was expected that the NN analysis would
exclude a greater region in the ($M_{Wbj},M_{bj}$)-plane than the cut-based
one, in fact neither excluded beyond $M_{\rho_T} \simeq 215\,\gev$ and
$M_{\pi_T} \simeq 120\,\gev$ for the default TCSM parameters used; see Fig.~3
in Ref.~\cite{Abazov:2006iq}.  Presumably, the NN analysis was limited by an
excess near that endpoint.

There are two recent CDF analyses, one in late 2006 based on
$955\,\ipb$~\cite{CDFa}, and a new one using
$1.9\,\ifb$~\cite{CDFb,Nagai:2008xq}. The TCSM parameters used were the same
as D\O's as far as the $\tro \ra W\tpi$ search is concerned. The 2008
analysis improves on the 2006 one, e.g., excluding $M_{\tpi} < 110\,\gev$,
$M_{\tro} < 210\,\gev$ at the 95\%~C.L. However, as with D\O, the CDF
exclusion plot (Fig.~4 in \cite{Nagai:2008xq}) falls well short of the
expected level at and above $M_{\tpi} = 125\,\gev$ and $M_{\tro} =
220\,\gev$.\footnote{Of course, if both experiments are seeing a signal,
  either D\O\ fluctuated up or CDF down.}

An interesting variable used in the CDF analyses is $Q = M_{Wbj} - M_{bj} -
M_W$. The resolution in $Q$ is a few GeV, much better than in $M_{bj}$ and
$M_{Wbj}$ alone, because jet energy uncertainties largely cancel in the
difference. If there is a narrow $\tro$ decaying to $W \tpi$, histograms of
$M_{bj}$ or $M_{Wbj}$ with $Q$ less than fixed values --- say 5, 10,
15$\dots$$50\,\gev$ --- will exhibit a sudden increase in one of the
invariant mass bins when, and only when, $Q$ has its resonant value.  In the
2008 analysis CDF presents color-coded two-dimensional plots of $M_{bj}$
vs.~$Q$-value. The colors in these plots are hard to distinguish; we urge CDF
to show histograms, with error bars, of $M_{bj}$ or $M_{Wbj}$ for a range of
cuts on $Q$.

In the remainder of this letter, we describe other studies and searches for
low-scale TC that may be possible at the Tevatron with larger data sets and
some that certainly can be carried out at the LHC. For our Tevatron
calculations, we use $M_{\tro} = M_{\tom} = 225\,\gev$, $M_{\ta} =
225$--$250\,\gev$ and $M_{\tpi} = 125\,\gev$. The isoscalar technipion
$\tpipr$ in the TCSM is assumed heavy, $300\,\gev$, so that none of the
lightest technivectors decay into it. The other TCSM parameters used here are
$\sin\chi = \third$, $Q_U = Q_D + 1 = 1$, $N_{TC} = 4$ and $M_{V_{1,2,3}} =
M_{A_{1,2,3}} = M_{\tro} = 225\,\gev$. The latter are defined in
Eqs.~(\ref{eq:amplitudes}) below.\footnote{The ALEPH Collaboration at LEP
  searched for a $\tro$ enhancement in $e^+ e^- \ra W_L^+ W_L^-$ and claimed
  a limit of $M_{\tro} > 600\,\gev$~\cite{Schael:2004tq}. The ALEPH analysis
  does not apply to the TCSM because it has a $\troz \ra W_L^+ W_L^-$
  coupling that is proportional to $\sin^2\chi \ll 1$. We have re-examined
  the ALEPH data and concluded that it sets {\em no} meaningful limit on the
  TCSM for the masses and other parameters assumed here. We shall present our
  analysis in a forthcoming paper.}  Typical cross sections for these
parameters are
\bea\label{eq:TeVxsects}
&&\sigma(\bar p p \ra \tropm \ra W^\pm \tpiz) \simeq 1.5\,\pb\,,
\quad \sigma(\bar p p \ra \troz \ra W^\pm \tpimp) \simeq 2.5\,\pb\,;\nn\\
&&\sigma(\bar p p \ra \tom \ra \gamma \tpiz) \simeq 0.3\,\pb\,,
\quad \sigma(\bar p p \ra \tom \ra \gamma Z^0) \simeq 0.07\,\pb\,;\nn\\
&&\sigma(\bar p p \ra \tom \ra \ellp\ellm) \simeq 0.2\,\pb\,,
\quad \sigma(\bar p p \ra \tom \ra \ellpm \nu_\ell) \simeq 0.3\,\pb\,.
\eea
The studies we propose include the angular distribution in $\tro \ra W \tpi$
and $WZ$; the decays $\tom \ra \gamma \tpi$, $\gamma Z$, $\ell^+\ell^-$ and
their angular distributions; and the search for $a_T^{\pm,0}$ and their decay
distributions. With even modest luminosity, the LHC should be able to
discover the $\tro$ and $\ta$ up to masses exceeding $500\,\gev$ and over a
wide range of TCSM parameters~\cite{Kreuzer:2006sd,Brooijmans:2008se}.
Discovering the $\tom$ at the LHC may require 10--$100\,\ifb$, depending on
its mass and decay mode.  We stress again that the studies carried out so far
to determine the LHC's reach for technicolor are preliminary and more careful
ones are needed. The LHC studies in
Refs.~\cite{Brooijmans:2008se,LSTCazuelos} used $M_{\tro} = M_{\tom} =
300$--$500\,\gev$, $M_{\ta} = 1.1 M_{\tro}$ and $M_{\tpi} = 200$--$350\,\gev$
(and $\sqrt{s} = 14\,\tev$).


{\em{2. {\underbar{The $\rho_T$ and the $\tom$}}}} The angular distribution
in $q \bar q \ra \tro \ra W\tpi$ is approximately $\sin^2\theta$, where
$\theta$ is the angle in the subprocess c.m.~frame between the incoming quark
and the outgoing $W$~\cite{Lane:2002sm}. The reason is is that 80--90\% of
this process is $q \bar q$ annihilation to $W_L \tpi$, effectively a pair of
pseudoscalars. Verification of this angular distribution would be strong
confirmation of its underlying TC origin. Requiring $Q \simeq M_{\rho_T} -
M_{\pi_T} - M_W$ will greatly enrich the signal-to-background for this
analysis, and the background angular distribution can be subtracted by
measuring it in sidebands.
 
 A second interesting study is the ratio of two-$b$-tag to one-$b$-tag events
 in $\tro \ra W \tpi$. The ratio of $\troz$ to $\tropm$ production is fairly
 well known because the relevant parton distribution functions are. The ETC
 coupling of $\tpi$ to quarks and leptons suggest that $\tpi$'s less massive
 than the top quark decay to $\bar b b$, $\bar b c$ and $\bar b u$. The
 two-$b$ to one-$b$ ratio tests this common but not theoretically
 well-established assumption. For example, Ref.~\cite{Zerwekh:2007pw}
 considers a search for $\tropm \ra \gamma \tpipm \ra \gamma \tau \nu_\tau$
 based on the supposition that $\tpipm \ra \bar b q$ are suppressed by
 CKM-like mixing angles.
 
 We hope that 4--$5\,\ifb$ at the Tevatron are sufficient to carry these
 studies out. Careful simulations of their signal and backgrounds are needed
 to determine that. The signal processes may be generated with {\sc
   Pythia}~\cite{Sjostrand:2006za}. The new release and its description may
 be found at {\tt www.hepforge.org}.
 
 At the LHC, the $\tro \ra W \tpi \ra \ell \nu bj$ signals are swamped by
 backgrounds from $t \bar t$ and $W$ plus heavy flavor production. The $t
 \bar t$ cross section is two orders of magnitude larger there than at the
 Tevatron! The best channel for a quick discovery and then observing the
 $\sin^2\theta$ distribution is $\tro^\pm \ra W^\pm Z^0 \ra \ell^\pm \nu_\ell
 \ell^+\ell^-$~\cite{Kreuzer:2006sd,Brooijmans:2008se}. For $e$ and/or $\mu$
 final states, the cross section times branching ratio ranges from about
 $100\,\fb$ for $M_{\tro} = 300\,\gev$ to $15\,\fb$ for $M_{\tro} =
 500\,\gev$. Integrated luminosities of $2.5$--$15\,\ifb$ are needed for
 $S/\sqrt{S+B} = 5\,\sigma$. The studies in Ref.~\cite{Brooijmans:2008se}
 indicate that the $\sin^2\theta$ distribution can be seen with
 (10,40,80)~$\ifb$ for $M_{\tro} =$ (300,400,500)~$\gev$.
 
 The $3\tpi$ decay channel of the $\tom$ is closed, while techni-isospin
 conservation greatly suppresses $\tom \ra W \tpi$. Therefore, for
 $M_{\omega_T} \simeq M_{\rho_T}$ and $Q_U + Q_D \simeq 1$, its major
 detectable decays at the Tevatron are $\tom \ra \gamma \tpiz \ra \gamma \bar
 b b$ and $\ell^+\ell^-$. For $Q_U + Q_D = 0$, $\tro$ and $\tom$ decays to
 $\gamma\tpiz$ are greatly suppressed and $\tom \ra \bar f f$ decays are
 forbidden altogether. It is very important that these final states are
 sought in the new high-luminosity data sets at the Tevatron.
 
 There is much to be done for the $\tom$ at the LHC. Its discovery may well
 be the first thing, and $\gamma Z^0$ {\em may be} the channel to focus on.
 The cross sections times branching ratios of $Z \ra e^+e^-$, $\mu^+\mu^-$
 are (20,6,3)~$\fb$ for $M_{\tom} =$ (300,400,500)~$\gev$.  The angular
 distributions for $\bar q q \ra \tom \ra \gamma \tpi$ and $\gamma Z_L^0$ are
 proportional to $1+\cos^2\theta$. The superb energy resolution achievable in
 the $\gamma Z \ra \gamma \ellp\ellm$ final states compensates somewhat for
 the lower signal rates~\cite{Brooijmans:2008se}, and should help distinguish
 the signal's $\cos\theta$ dependence from the background's. We also urge
 that $\tom \ra \ellp\ellm$ studies be carried out. These depend on TCSM
 parameters and may help determine them.
 

 
 {\em{3. {\underbar{The $a_T$}}}} If $\ta$ and $\tro$ are close enough in
 mass to reduce the $S$-parameter to an acceptable level, the channels $\ta
 \ra 3\tpi$ and $\tro\tpi$ certainly are closed. The two-body decay modes
 consistent with techni-isospin symmetry (for ``strong'' decays) and CP
 conservation are then $\ta \ra G \tpi$ and $G W_L$, $G V_T$, and $W_L V_T$
 where $V_T = \tro$ or $\tom$, $G = \gamma$ or a transversely polarized $W$
 or $Z$. Following Refs.~\cite{Lane:1999uh,Lane:2002sm}, the decay amplitudes
 are:
\bea\label{eq:amplitudes}
&&\CM(a_{T}(p_1) \ra G(p_2) \pi_{T}(p_3)) = \frac{eV_{a_T G_A \pi_T}}{2M_{V_2}}
\,\widetilde F_1^{\lambda\mu}F^*_{2\lambda\mu} + \frac{eA_{a_T G_V
    \pi_T}}{2M_{A_2}}
\,F_1^{\lambda\mu}F^*_{2\lambda\mu}\,;\\
&& \CM(a_{T}(p_1) \ra G(p_2) V_{T}(p_3)) = \frac{eV_{a_T G_V V_T}}{2M^2_{V_3}}
\,\widetilde F_1^{\lambda\mu}F^*_{2\mu\nu}F^{*\nu}_{3\lambda}
 + \frac{eA_{a_T G_A V_T}}{2M^2_{A_3}}
 F_1^{\lambda\mu}F^*_{2\mu\nu}F^{*\nu}_{3\lambda}\,;\\
&&\CM(a_{Ti}(p_1) \ra  \rho_{Tj}(p_2) W_{Lk}(p_3)) =
\frac{g_{a_T\rho_T\pi_T}}{2M_{a_T}}\,\epsilon_{ijk}\,\sin\chi\,
F_1^{\lambda\mu}F^*_{2\lambda\mu}\,.
\eea
\begin{figure}[!t]
  \begin{center}
    \includegraphics[width=3.10in, height = 4.00in, angle=90, trim=60 60 60 60]
    {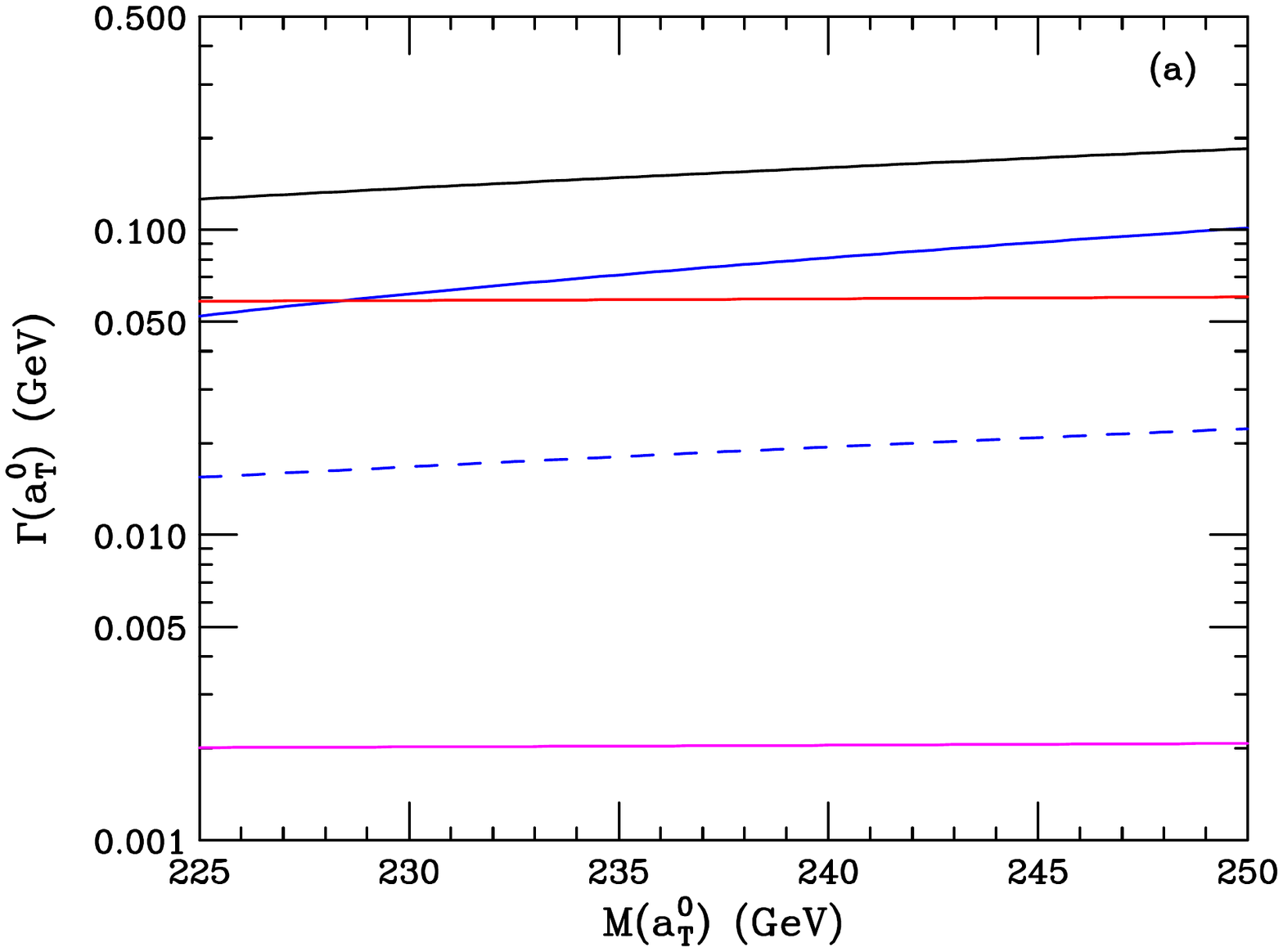}
    \includegraphics[width=3.10in, height = 4.00in, angle=90, trim=60 60 60 60]
    {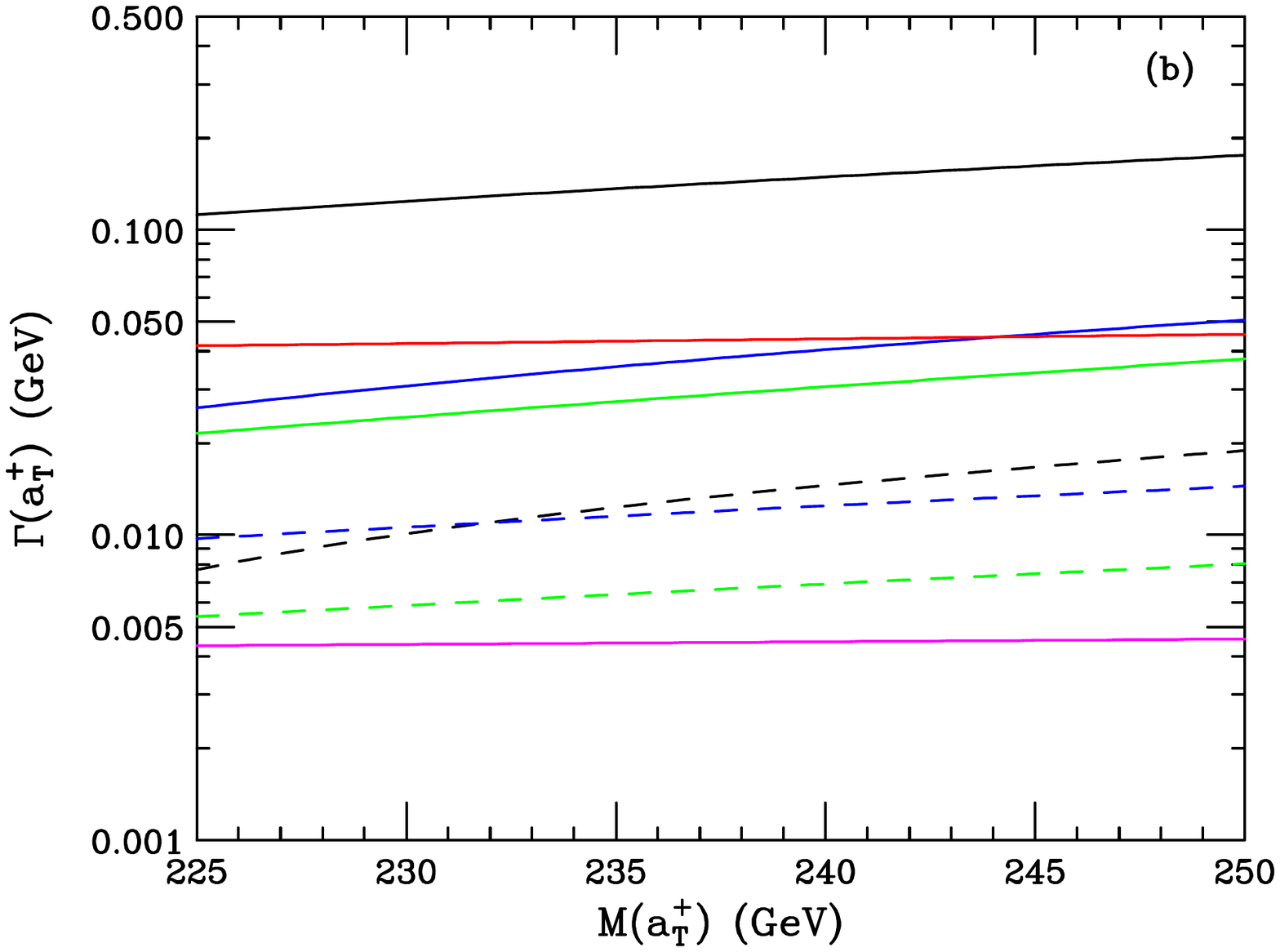}
     \caption{$a_T$ Decay rates for parameters given in the text. Total width
       (black). (a) $a_T^0 \ra W^\pm\tpimp$ (blue), $\sum_i \bar f_i f_i$
       (red), $W^+W^-$ (blue dashed) and $\ellp\ellm$ (magenta). (b) $a_T^\pm
       \ra W^\pm\tpiz$ (blue), $\gamma \tpipm$ (green), $\sum_i \bar
       f^{\prime}_i f_i$ (red), $Z^0\tpipm$ (black dashed), $W^\pm Z^0$ (blue
       dashed), $\gamma W^\pm$ (green dashed) and $\ellpm \nu_\ell$
       (magenta).}
    \label{LSTC_fig_1}
  \end{center}
\end{figure}
Phase-space limitations imply that only $G\tpi$ and $G W_L$ are important.
The $a_T$ also decays to fermion pairs via the $W$ and $Z$; these modes
become dominant (but do not lead to large cross sections at the Tevatron)
when the $M_{V,A}$ are large. In Eqs.~(3)--(5), $F_{n\lambda\mu} =
\epsilon_{n\lambda} \, p_{n\mu} - \epsilon_{n\mu} \, p_{n\lambda}$ and
$\widetilde F_{n\lambda\mu} = \half \epsilon_{\lambda\mu\nu\rho}
F_n^{\nu\rho}$; $(i,j,k)$ are isospin indices.  The TCSM mass parameters
$M_{V_{2,3}}$, $M_{A_{2,3}}$ are similar to $M_{V,A} \equiv M_{V_1,A_1}$ in
Ref.~\cite{Lane:2002sm} and are expected to be $\CO(M_{\tro})$. The factors
$V_{a_T G_A \pi_T} = 2\,{\rm Tr}(Q_{a_T}\{Q_{G_A}^\dagg, Q_{\pi_T}^\dagg\})$
and $A_{a_T G_V \pi_T} = 2\,{\rm Tr}(Q_{a_T}[Q_{G_V}^\dagg,
Q_{\pi_T}^\dagg])$ are given in Ref.~\cite{Lane:2002sm} by using $Q_{a_T} =
Q_{\rho_T}$ and the other charges as defined there. The couplings of $a_T$ to
the axial part of the weak bosons $G_A = (W_A,\, Z_A$) are $f_{Ga_T} = 2
\sqrt{\alpha/\alpha_{a_T}} {\rm Tr}(Q_{a_T}Q^\dagger_{G_A}) =
(-\sqrt{\alpha/\alpha_{a_T}}/(2\sin\thw),\,
-\sqrt{\alpha/\alpha_{a_T}}/\sin(2\thw))$. These enter the bosons' propagator
matrices. Here, $\alpha_{a_T} = g^2_{a_T}/4\pi$ is analogous to
$\alpha_{\rho_T}$ of the TCSM and is defined by $\langle\Omega|\half \bar
T\gamma_\mu\gamma_5 \tau_i T |a_{Tj}(p)\rangle = M_{a_T}^2 \epsilon_\mu(p)\,
\delta_{ij}/g_{a_T}$. If we scale $g_{\rho_T}$ from the $\rho\pi\pi$ coupling
$g_{\rho}$ determined from the decay $\tau \ra \rho\nu_\tau$, and set
$g_{a_T} = g_{\rho_T}$ to make $S$ small, then $\alpha_{a_T} =
2.16\,(3/N_{TC})$ in the TCSM.

 Sample decay rates are shown in Fig.~\ref{LSTC_fig_1} for $225 < M_{\ta} <
 250\,\gev$ and the other TCSM parameters we used in our Tevatron
 calclations. The most important decay modes are $a_T^0 \ra W^\pm \pi_T^\mp$,
 $W^+W^-$, and $\ellp\ellm$; $a_T^+ \ra W^+ \tpiz$, $\gamma\tpip$, $Z^0
 \tpip$, and $W^+Z^0$ and, perhaps, $\gamma W^+$. These decays yield
 prominent signatures involving leptons, missing transverse energy, photons
 and $b$-jets. As for $\tro$ and $\tom$, the $\ta$ are very narrow.

 \begin{figure}[!t]
   \begin{center}
     \vspace*{-0.5in}
     \includegraphics[width=3.10in, height = 4.00in, angle=90, trim=60 60 60 60]
     {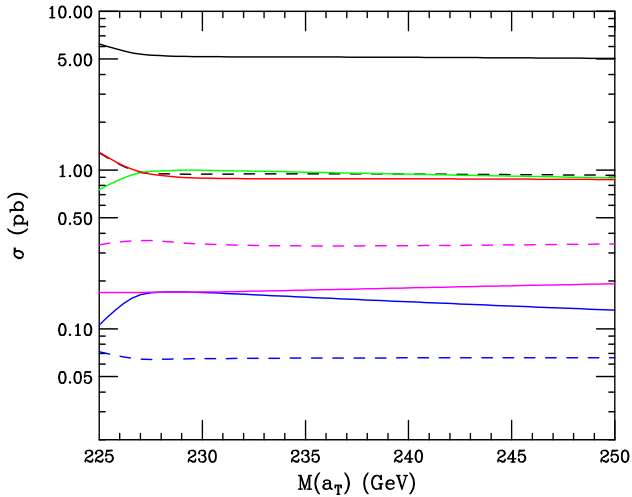}
     \caption{Production rates for $W\tpi$ (black), $Z\tpi$ (red),
       $\gamma\tpi$ (green) $WZ$ (black dashed), $\gamma W$ (blue), $\gamma
       Z$ (blue dashed), $\ellp\ellm$ (magenta) and $\ellpm\nu_\ell$
       (magenta dashed) in $\bar p p$ collisions at the Tevatron with
       $\sqrt{s} = 1.96\,\tev$. All charge modes are summed. $M_{\tro} =
       M_{\tom} = 225\,\gev$ and $M_{\tpi} = 125\,\gev$, and other parameters
       are given in the text.}
     \label{LSTC_fig_2}
   \end{center}
 \end{figure}

 In Fig.~\ref{LSTC_fig_2} we display the signal production rates at the
 Tevatron for $W^{\pm}\pi_T^{\mp,0}$, $\gamma\pi_T^{\pm,0}$ (summed over all
 charges), $W^\pm Z^0$, $Z^0\tpipm$, $\gamma W^\pm$, $\gamma Z^0$,
 $\ellp\ellm$ and $\ellpm\nu_\ell$.\footnote{No K-factor has been applied to
   these cross section estimates. Standard-model contributions to $WZ$,
   $\gamma W$ and $\gamma Z$ rates are not included.  For narrow resonances
   at a hadron collider, they may be added incoherently to the signal rates.}
 These include contributions from $\tro$, $\tom$ and $\at$ intermediate
 states. The cross sections without the $a_T$ present are 3.9, 0.45, 0.68,
 0.67, 0.04, 0.07, 0.13 and $0.20\,\pb$, respectively. So, e.g., there is
 about $1\,\pb$ of $a_T^{\pm,0} \ra W \tpi$ and $0.3\,\pb$ of $\tapm \ra
 \gamma \tpipm$. As $M_{V_i,A_i}$ is increased, the rates for processes
 involving a transverse EW boson decrease while $\ellp\ellm$ and
 $\ellpm\nu_\ell$ rates increase. In hadron colliders, all the production
 comes from these narrow resonances, not their tails, so that the invariant
 masses of the final states are sharply defined. If $M_{\ta} \simge
 M_{\tro,\,\tom} + 20\,\gev$, it should be possible to discern the separate
 vector and axial vector contributions. The more precisely measured $Q$-value
 will be helpful.\footnote{Just such a study was carried out for the ATLAS
   detector at the LHC, with encouraging results, for $\tropm, \tapm \ra Z^0
   \tpipm \ra \ellp\ellm b \bar q$ in Ref.~\cite{LSTCazuelos}.}  Also,
 determination of the final state angular distribution may clarify what is
 happening.
 
 Similar rates as in Fig.~\ref{LSTC_fig_2} occur at the LHC for technihadron
 masses that are 1.5--2 times as large. As already mentioned, $\tropm \ra
 W^\pm Z$ and $\tom \ra \gamma Z$ (and possibly $\gamma \tpi$) have the most
 manageable backgrounds. The $\tapm$ is best sought in its $\gamma W_L^\pm$
 mode. The cross section times branching ratio to $e$ and $\mu$ final states
 is (170,65,30)~$\fb$ for $M_{\ta} =$ (330,440,550)~$\gev$. As for $\tom \ra
 \gamma Z_L$, this mode has a $1 + \cos^2\theta$ angular distributions. The
 best hope for discovering $\taz$ at the LHC appears to be the $\ellp\ellm$
 channel.
 
 In conclusion, the time is ripe for dedicated searches for low-scale
 technicolor at the Tevatron. There is, or soon will be, enough data to
 discover, or rule out, $\tro$ and $\tpi$ with masses below about $250\,\gev$
 and $150\,\gev$. If $\tro \ra W\tpi$ is found, its decay angular
 distribution is approximately $\sin^2\theta$, an important confirmation of
 the underlying technicolor dynamics. It will also be profoundly important to
 search for $\tom \ra \gamma\tpi \ra \gamma \bar b b$. There is good
 theoretical reason to expect that the spectrum of low-scale technicolor is
 richer than heretofore thought, with the axial vector state $\at$
 approximately degenerate with its $\tro$ partner. The axial states also
 decay into EW gauge bosons plus a technipion. At the LHC, the most promising
 modes appear to be $\tropm \ra W_L^\pm Z_L^0$, $\tom \ra \gamma Z_L^0$ and
 $\tapm \ra \gamma W_L^\pm$. The first has a $\sin^2\theta$ decay
 distribution while the other two are $1 + \cos^2\theta$. Luminosities of a
 few to a few 10s of $\ifb$ are sufficient to discover $\tro$ and $\ta$ up to
 about $500\,\gev$; the $\tom$ may require 10--$100\,\ifb$. Finally,
 $\tom,\taz \ra \ellp\ellm$ are likely to be the most promising ways to study
 these states at the LHC. Detailed simulations are needed. We urge the
 detector collaborations to carry them out.

 
 {\em{\underbar{Acknowledgments}}} We are especially grateful to Meenakshi
 Narain and Weiming Yao for discussions about the search for technihadrons
 and to Steve Mrenna for including the new $\ta$ processes in {\sc Pythia}.
 We thank Adam Martin and Veronica Sanz for many valuable conversations. We
 also thank Michael Barnett, Ken Hayes, and Colin Morningstar for helpful
 input on a technical point. K.L.~also thanks Laboratoire d'Annecy-le-Vieux
 de Physique Theorique for its hospitality and support. FNAL is operated by
 Fermi Research Alliance, LLC, under contract DE-AC02-07CH11359 with the
 U.S.~Department of Energy.  KL's research is supported by the Department of
 Energy under Grant~No.~DE-FG02-91ER40676

\vfil\eject

\bibliography{LSTC}

\providecommand{\href}[2]{#2}\begingroup\raggedright\begin{thebibliography}{10}

\bibitem{Holdom:1981rm}
B.~Holdom, ``Raising the sideways scale,'' {\em Phys. Rev.} {\bf D24} (1981)
  1441.

\bibitem{Appelquist:1986an}
T.~W. Appelquist, D.~Karabali, and L.~C.~R. Wijewardhana, ``Chiral hierarchies
  and the flavor changing neutral current problem in technicolor,'' {\em Phys.
  Rev. Lett.} {\bf 57} (1986) 957.

\bibitem{Yamawaki:1986zg}
K.~Yamawaki, M.~Bando, and K.-i. Matumoto, ``Scale invariant technicolor model
  and a technidilaton,'' {\em Phys. Rev. Lett.} {\bf 56} (1986) 1335.

\bibitem{Akiba:1986rr}
T.~Akiba and T.~Yanagida, ``Hierarchic chiral condensate,'' {\em Phys. Lett.}
  {\bf B169} (1986) 432.

\bibitem{Eichten:1984eu}
E.~Eichten, I.~Hinchliffe, K.~D. Lane, and C.~Quigg, ``{Super Collider
  Physics},'' {\em Rev. Mod. Phys.} {\bf 56} (1984) 579--707.

\bibitem{Lane:1989ej}
K.~D. Lane and E.~Eichten, ``Two scale technicolor,'' {\em Phys. Lett.} {\bf
  B222} (1989) 274.

\bibitem{Eichten:1996dx}
E.~Eichten and K.~D. Lane, ``Low-scale technicolor at the Tevatron,'' {\em
  Phys. Lett.} {\bf B388} (1996) 803--807,
  \href{http://xxx.lanl.gov/abs/hep-ph/9607213}{ hep-ph/9607213}.

\bibitem{Eichten:1997yq}
E.~Eichten, K.~D. Lane, and J.~Womersley, ``Finding low-scale technicolor at
  hadron colliders,'' {\em Phys. Lett.} {\bf B405} (1997) 305--311,
  \href{http://xxx.lanl.gov/abs/hep-ph/9704455}{ hep-ph/9704455}.

\bibitem{Eichten:1979ah}
E.~Eichten and K.~D. Lane, ``Dynamical Breaking of Weak Interaction
  Symmetries,'' {\em Phys. Lett.} {\bf B90} (1980) 125--130.

\bibitem{Lane:1999uh}
K.~D. Lane, ``Technihadron production and decay in low-scale technicolor,''
  {\em Phys. Rev.} {\bf D60} (1999) 075007,
  \href{http://xxx.lanl.gov/abs/hep-ph/9903369}{ hep-ph/9903369}.

\bibitem{Lane:2002sm}
K.~Lane and S.~Mrenna, ``The collider phenomenology of technihadrons in the
  technicolor Straw Man Model,'' {\em Phys. Rev.} {\bf D67} (2003) 115011,
  \href{http://xxx.lanl.gov/abs/hep-ph/0210299}{ hep-ph/0210299}.

\bibitem{Hill:1994hp}
C.~T. Hill, ``Topcolor assisted technicolor,'' {\em Phys. Lett.} {\bf B345}
  (1995) 483--489, \href{http://xxx.lanl.gov/abs/hep-ph/9411426}{
  hep-ph/9411426}.

\bibitem{Lane:1993wz}
K.~D. Lane, ``An Introduction to technicolor,''
  \href{http://xxx.lanl.gov/abs/hep-ph/9401324}{ hep-ph/9401324}.

\bibitem{Lane:1994pg}
K.~D. Lane, ``Technicolor and precision tests of the electroweak
  interactions,'' \href{http://xxx.lanl.gov/abs/hep-ph/9409304}{
  hep-ph/9409304}.

\bibitem{Peskin:1990zt}
M.~E. Peskin and T.~Takeuchi, ``A new constraint on a strongly interacting
  Higgs sector,'' {\em Phys. Rev. Lett.} {\bf 65} (1990) 964--967.

\bibitem{Golden:1990ig}
M.~Golden and L.~Randall, ``Radiative corrections to electroweak parameters in
  technicolor theories,'' {\em Nucl. Phys.} {\bf B361} (1991) 3--23.

\bibitem{Holdom:1990tc}
B.~Holdom and J.~Terning, ``Large corrections to electroweak parameters in
  technicolor theories,'' {\em Phys. Lett.} {\bf B247} (1990) 88--92.

\bibitem{Altarelli:1991fk}
G.~Altarelli, R.~Barbieri, and S.~Jadach, ``Toward a model independent analysis
  of electroweak data,'' {\em Nucl. Phys.} {\bf B369} (1992) 3--32.

\bibitem{Appelquist:1998xf}
T.~Appelquist and F.~Sannino, ``The physical spectrum of conformal SU(N) gauge
  theories,'' {\em Phys. Rev.} {\bf D59} (1999) 067702,
  \href{http://xxx.lanl.gov/abs/hep-ph/9806409}{ hep-ph/9806409}.

\bibitem{Knecht:1997ts}
M.~Knecht and E.~de~Rafael, ``Patterns of spontaneous chiral symmetry breaking
  in the large N(c) limit of QCD-like theories,'' {\em Phys. Lett.} {\bf B424}
  (1998) 335--342, \href{http://xxx.lanl.gov/abs/hep-ph/9712457}{
  hep-ph/9712457}.

\bibitem{Appelquist:1999dq}
T.~Appelquist, P.~S. Rodrigues~da Silva, and F.~Sannino, ``{Enhanced global
  symmetries and the chiral phase transition},'' {\em Phys. Rev.} {\bf D60}
  (1999) 116007, \href{http://xxx.lanl.gov/abs/hep-ph/9906555}{
  hep-ph/9906555}.

\bibitem{Hirn:2006nt}
J.~Hirn and V.~Sanz, ``A negative S parameter from holographic technicolor,''
  {\em Phys. Rev. Lett.} {\bf 97} (2006) 121803,
  \href{http://xxx.lanl.gov/abs/hep-ph/0606086}{ hep-ph/0606086}.

\bibitem{Hirn:2006wg}
J.~Hirn and V.~Sanz, ``The fifth dimension as an analogue computer for strong
  interactions at the LHC,'' \href{http://xxx.lanl.gov/abs/hep-ph/0612239}{
  hep-ph/0612239}.

\bibitem{Kreuzer:2006sd}
P.~Kreuzer, ``{Search for technicolor at CMS in the rho(TC) $\to$ W + Z
  channel},''. CERN-CMS-NOTE-2006-135.

\bibitem{Kreuzer:2007zz}
P.~Kreuzer, ``{Technicolour and other beyond the standard model alternatives in
  CMS},'' {\em Acta Phys. Polon.} {\bf B38} (2007) 459--468.

\bibitem{Brooijmans:2008se}
G.~Brooijmans {\em et.~al.}, ``{New Physics at the LHC: A Les Houches Report.
  Physics at Tev Colliders 2007 -- New Physics Working Group},''
  \href{http://xxx.lanl.gov/abs/0802.3715}{ 0802.3715}.

\bibitem{LSTCazuelos}
G.~Azuelos, J.~Ferland, K.~Lane, and A.~Martin, ``Search for Low-Scale
  Technicolor in ATLAS.'' ATLAS Note, ATL-PHYS-CONF-2008-003, 2008.

\bibitem{Abazov:2006iq}
{\bf D0} Collaboration, V.~M. Abazov {\em et.~al.}, ``{Search for
  techniparticles in e + jets events at D0},'' {\em Phys. Rev. Lett.} {\bf 98}
  (2007) 221801, \href{http://xxx.lanl.gov/abs/hep-ex/0612013}{
  hep-ex/0612013}.

\bibitem{CDFa}
{\bf CDF} Collaboration, ``Search for Technicolor Particles Produced in
  Association with W Bosons at CDF.''
  http://www-cdf.fnal.gov/physics/exotic/r2a/20061025.techcolor/.

\bibitem{CDFb}
{\bf CDF} Collaboration, ``Search for Technicolor Particles Produced in
  Association with $W^\pm$ Boson with 1.9 fb$^{-1}$ at CDF.''
  http://www-cdf.fnal.gov/physics/new/hdg/results/technicolor-080411/.

\bibitem{Nagai:2008xq}
Y.~Nagai, T.~Masubuchi, S.~Kim, and W.~M. Yao, ``{Search for Technicolor
  Particles Produced in Association with W Boson at CDF},''
  \href{http://xxx.lanl.gov/abs/0808.0226}{ 0808.0226}.

\bibitem{Schael:2004tq}
{\bf ALEPH} Collaboration, S.~Schael {\em et.~al.}, ``{Improved measurement of
  the triple gauge-boson couplings gamma W W and Z W W in e+ e- collisions},''
  {\em Phys. Lett.} {\bf B614} (2005) 7--26.

\bibitem{Zerwekh:2007pw}
A.~R. Zerwekh, C.~O. Dib, and R.~Rosenfeld, ``Technicolor contribution to
  lepton + photon + missing E(T) events at the Tevatron,''
  \href{http://xxx.lanl.gov/abs/hep-ph/0702167}{ hep-ph/0702167}.

\bibitem{Sjostrand:2006za}
T.~Sjostrand, S.~Mrenna, and P.~Skands, ``PYTHIA 6.4 physics and manual,'' {\em
  JHEP} {\bf 05} (2006) 026, \href{http://xxx.lanl.gov/abs/hep-ph/0603175}{
  hep-ph/0603175}.

\end{thebibliography}\endgroup
\bibliographystyle{utcaps}
\end{document}